%% file: main.tex
\documentclass[onecolumn]{aastex63}
\usepackage{natbib}
\begin{document}

\shorttitle{Systematics}
\shortauthors{Carpenter et al.}

\title{Update on the Systematics in the ALMA Proposal Review Process after Cycle 8}

\author[0000-0003-2251-0602]{John M. Carpenter}
\affiliation{Joint ALMA Observatory, Avenida Alonso de C\'ordova 3107, Vitacura, Santiago, Chile}

\author{Andrea Corvill\'on}
\affiliation{Joint ALMA Observatory, Avenida Alonso de C\'ordova 3107, Vitacura, Santiago, Chile}

\author[0000-0002-3106-7676]{Jennifer Donovan Meyer}
\affiliation{National Radio Astronomy Observatory (NRAO), 520 Edgemont Road, Charlottesville, VA 22903, USA}

\author[0000-0002-9912-5705]{Adele L. Plunkett}
\affiliation{National Radio Astronomy Observatory (NRAO), 520 Edgemont Road, Charlottesville, VA 22903, USA}

\author{Robert Kurowski}
\affiliation{ESO Garching, Karl-Schwarzschild-Str. 2, 85748, Garching bei Munchen, Germany}

\author{Alex Chalevin}
\affiliation{ESO Garching, Karl-Schwarzschild-Str. 2, 85748, Garching bei Munchen, Germany}

\author[0000-0003-1283-6262]{Enrique Mac\'ias}
\affiliation{Joint ALMA Observatory, Avenida Alonso de C\'ordova 3107, Vitacura, Santiago, Chile}
\affiliation{ESO Chile, Avenida Alonso de C\'ordova 3107, Vitacura, Santiago, Chile}
\affiliation{ESO Garching, Karl-Schwarzschild-Str. 2, 85748, Garching bei Munchen, Germany}

\correspondingauthor{John M. Carpenter}
\email{john.carpenter@alma.cl}

\input{abstract}

\input{introduction}
\input{overview}
\input{systematics}
\input{conclusions}

\clearpage

\bibliography{references}

\end{document}

%% file: abstract.tex
\begin{abstract}

We present an updated analysis of systematics in the Atacama Large Millimeter/submillimeter Array (ALMA) proposal ranks from \citet{Carpenter20a} to include the last two ALMA cycles, when significant changes were introduced in the proposal review process. In Cycle 7, the investigator list on the proposal cover sheet was randomized such that the reviewers were aware of the overall proposal team but did not know  the identity of the principal investigator (PI). In Cycle 8, ALMA adopted distributed peer review for most proposals and implemented dual-anonymous review for all proposals, in which the identity of the proposal team was not revealed to the reviewers. The most significant change in the systematics in Cycles 7 and 8 compared to previous cycles is related to the experience of PIs in submitting ALMA proposals. PIs that submit a proposal every cycle tend to have ranks that are consistent with average in Cycles 7 and 8 whereas previously they had the best overall ranks. Also, PIs who submitted a proposal for the second time show improved ranks over previous cycles. These results suggest some biases related to the relative prominence of the PI have been present in the ALMA review process. Systematics related to regional affiliation remain largely unchanged in that PIs from Chile, East Asia, and non-ALMA regions tend to have poorer overall ranks than PIs from Europe and North America. The systematics of how one region ranks proposals from another region are also investigated. No significant differences in the overall ranks based on gender of the PI are observed.

\end{abstract}

%% file: introduction.tex
\section{Introduction}
\label{sec:intro}

The proposal review process is a critical function of any astronomical observatory. The selected proposals set the science objectives for the near future, and the success of a proposal is important for the scientific development of the proposal teams. It is obviously imperative that the process is as fair and impartial as possible.

\citet{Reid14} brought attention to potential biases in the proposal review process of the Hubble Space Telescope (HST) when he found that female principal investigators (PIs) have had a lower acceptance rate of proposals than male PIs. The difference was not formally significant in any given cycle, but was present and persistent for over a decade. Since that study, HST has adopted a number of measures to address these potential biases, including recently introducing dual-anonymous review where the identity of the proposal team is not revealed to the reviewers. While the impact of these changes needs to be evaluated over several cycles, \citet{Johnson20} found that HST proposals from female PIs have been more likely to be accepted since HST began making changes to the proposal review process to address the gender inequality. In addition, with the implementation of dual-anonymous review, the percentage of accepted proposals awarded to PIs that had not been previously successful on HST increased to 30\%, up from 5\% in the prior decade \citep{SinghChawla21}.

Other observatories have followed the lead of HST and have begun to publish their statistics related to potential systematics in their review processes, especially pertaining to gender. \citet{Patat16} found a lower proposal acceptance rate for women compared to men for proposals submitted to the European Southern Observatory (ESO) in their bi-annual proposal calls. Some of the differences may be explained by demographics in that senior PIs, of which a higher percentage are men, also have higher proposal acceptance rates than junior researchers. However, even after removing trends based on seniority, women have a lower acceptance rate than men. \citet{Lonsdale16} analyzed the results from the proposal review process for facilities operated in full or in part by the National Radio Astronomy Observatory (NRAO), including the Jansky Very Large Array (JVLA), the Very Long Baseline Array (VLBA), the Green Bank Telescope (GBT) and the Atacama Large Millimeter/submillimeter Array (ALMA). Similar to HST and ESO, they found that men tend to have better proposal ranks than women in ALMA Cycles 2-4. Similar trends were found for the JVLA, VLBA, and GBT, although in some semesters, women had better overall rankings than men. Since the \citet{Lonsdale16} study, NRAO has taken steps to reduce biases for the JVLA, VLBA, and GBT by informing the review panels and time allocation committee about gender imbalances and increasing the female representation in the review process. No significant differences in the rankings between men and women have been found since 2018 \citep{Hunt21}. 

\citet{Carpenter20a} extended the results from \citet{Lonsdale16} pertaining to ALMA by including more ALMA cycles and by investigating potential systematics related to the regional affiliation of the PI and the experience level of the PI in submitting ALMA proposals. \citet{Carpenter20a} showed that while the differences in the rankings between men and women have diminished since ALMA Cycles 2-4, in each cycle through Cycle 6 women have had a lower proposal acceptance rate than men even after differences in demographics are considered.
Other systematics were found in that PIs from Europe and North America have better proposal ranks than PIs from Chile, East Asia, and other regions. The origin of the systematics by region is unclear, but \citet{Carpenter20a} speculated that it could be related to English being a non-native language for the majority of PIs in those regions, or differences in communication styles by region. It was also found that PIs who submit proposals every single cycle generally have the best overall proposal ranks, while PIs who have submitted proposals for the first or second time generally have the poorest ranks.

Since the study by \citet{Carpenter20a}, ALMA has made changes to the proposal review process in an attempt to reduce biases. The main change introduced in 2019 (Cycle 7) was that the investigator list was randomized on the proposal cover sheet and the first names were listed with the first initial only, following an earlier practice by HST \citep{Johnson20}.
In 2021 (Cycle 8), ALMA adopted a dual-anonymous review where the investigator team is not revealed to the reviewers. Also in Cycle 8, ALMA adopted a new review process for the majority of proposals called distributed peer review \citep{Merrifield09}, where each PI designates one member of the proposal team to review a small number of proposals. While distributed peer review was not implemented to address systematics in the rankings, the impact, if any, of this new process on systematics needs careful examination.

In this paper, we update the analysis presented in \citet{Carpenter20a} and investigate any changes in the systematics in the past two ALMA cycles (Cycle 7 and Cycle 8). Section~\ref{sec:overview} provides an overview of the ALMA proposal review process. Section~\ref{sec:analysis} analyzes the proposal rankings to identify any changes in the systematics in the past two cycles. Section~\ref{sec:conclusions} summarizes the results and conclusions of this study.

%% file: overview.tex
\section{The ALMA Proposal Review Process}
\label{sec:overview}

\citet{Carpenter20a} described the ALMA proposal review process as applied in ALMA Cycles 0-6. Briefly, the review process consisted of review panels dedicated to specific topics, where the Joint ALMA Observatory (JAO) invited members of the community to serve on the panels. In Stage 1, reviewers scored the proposals and wrote preliminary comments. The Stage 1 scores were averaged to create a preliminary ranked list. The poorest ranked proposals from Stage 1 were dropped from further consideration in a ``triage" process. The individual panels then met to discuss the proposals assigned to their panels and to rescore the non-triaged proposals. The final scores determined the overall ranked list of proposals, which is the primary output of the review panels.

The ALMA Cycle 7 proposal review process followed the same basic process as in the previous cycles except that the number of panels was increased to 25 from 18 to accommodate the large number of submitted proposals. Also, to reduce potential biases, the list of investigators on the proposal cover sheet was randomized such that the reviewers were aware of the overall proposal team, but the PI was no longer identified. In addition, the list of investigators was removed from the software tools used by reviewers. In Cycle 8, ALMA adopted a dual-anonymous policy such that the identity of the proposal team was not provided to the reviewers. Accordingly, PIs were required to write their proposals in a manner that does not reveal the proposal team, primarily by not referencing their own research in the first person. 

ALMA also introduced a significant structural change to the review process in Cycle 8 with distributed peer review. In distributed peer review, each proposal team nominates one member of the proposal team to participate in the review process. As applied by ALMA, each reviewer is assigned 10 proposals, which they rank in priority order from 1 (strongest) to 10 (weakest). The individual ranks from the reviewers for a given proposal are averaged to derive the overall proposal rank. Distributed peer review also has a second stage where reviewers can read the comments from other reviewers, and they have the option to modify their own ranks and reviews. The review is conducted entirely remotely with no face-to-face discussion. Donovan Meyer et al. (2022, submitted) describes the Cycle 8 distributed peer review process in detail and presents an analysis of the results. ALMA previously used this review method in the Cycle 7 Supplemental Call for Proposals to assess 249 proposals for the 7-m Array, the results of which are presented in \citet{Carpenter20b}. The overall process adopted by ALMA broadly follows that introduced by \citet{Merrifield09}, but with one significant exception. \citet{Merrifield09} proposed an incentive system to reward ``good" refereeing and discourage ``bad" refereeing by modifying the position of the reviewer's own proposal in the final ranked list depending on the degree to which the reviewer's ranks agreed with the overall consensus. ALMA did not adopt such an approach on the belief that the variations in assigned ranks between reviewers is largely based on true differences of opinion regarding the relative merit of proposals.

The Cycle 8 Main Call for Proposals used distributed peer review to assess the 1497 submitted proposals that requested less than 25 h on the 12-m Array or less than 150 h on the 7-m Array in ``standalone" mode. An additional 238 proposals that requested more than 25 h on the 12-m Array, including 40 Large Programs, were reviewed by topical panels that followed the same basic procedure used in previous cycles. The analysis presented here is for the combined results of distributed peer review and the review panels. For the proposals reviewed by the panels, the Stage 1 scores were used to determine the overall ranks because not all proposals are reviewed in Stage 2 as a result of triage, and because the triage process was applied depending on the oversubscription rate in a given region. Since more proposals in a heavily oversubscribed region (e.g., Europe) were triaged than proposals in a less subscribed region (e.g., Chile), biases with respect to region will be introduced in the Stage 2 rankings. As shown in \citet{Carpenter20a}, any systematics in the rankings are introduced mainly in the Stage 1 process. The overall rankings from distributed peer review were normalized by the number of proposals in the distributed peer review process. Similarly, the ranks within a given panel were normalized by the number of proposals in that panel. The normalized ranks from distributed peer review and the panels were combined and sorted to determine the combined ranked list of proposals.

%% file: systematics.tex
\section{Analysis of the proposal rankings}
\label{sec:analysis}

Following \citet{Carpenter20a}, we investigated the presence of systematics based on the experience of the PI in submitting ALMA proposals, the regional affiliation of the PI, and the gender of the PI.

\subsection{Experience level}

Experience is defined as the number of cycles in which a PI has submitted a proposal in the Main Call. This metric provides a measure of how familiar a PI is in preparing ALMA proposals, but it does not necessarily reflect the career status of the PI or their expertise in interferometry. \citet{Carpenter20a} showed  that the most experienced PIs (i.e., those that submit a proposal in every cycle) have had the best overall ranks in each of Cycles 1 to 6. Conversely, the first-time PIs have had the poorest overall ranks in each cycle.

\begin{figure}
\centering
\includegraphics[width=\textwidth]{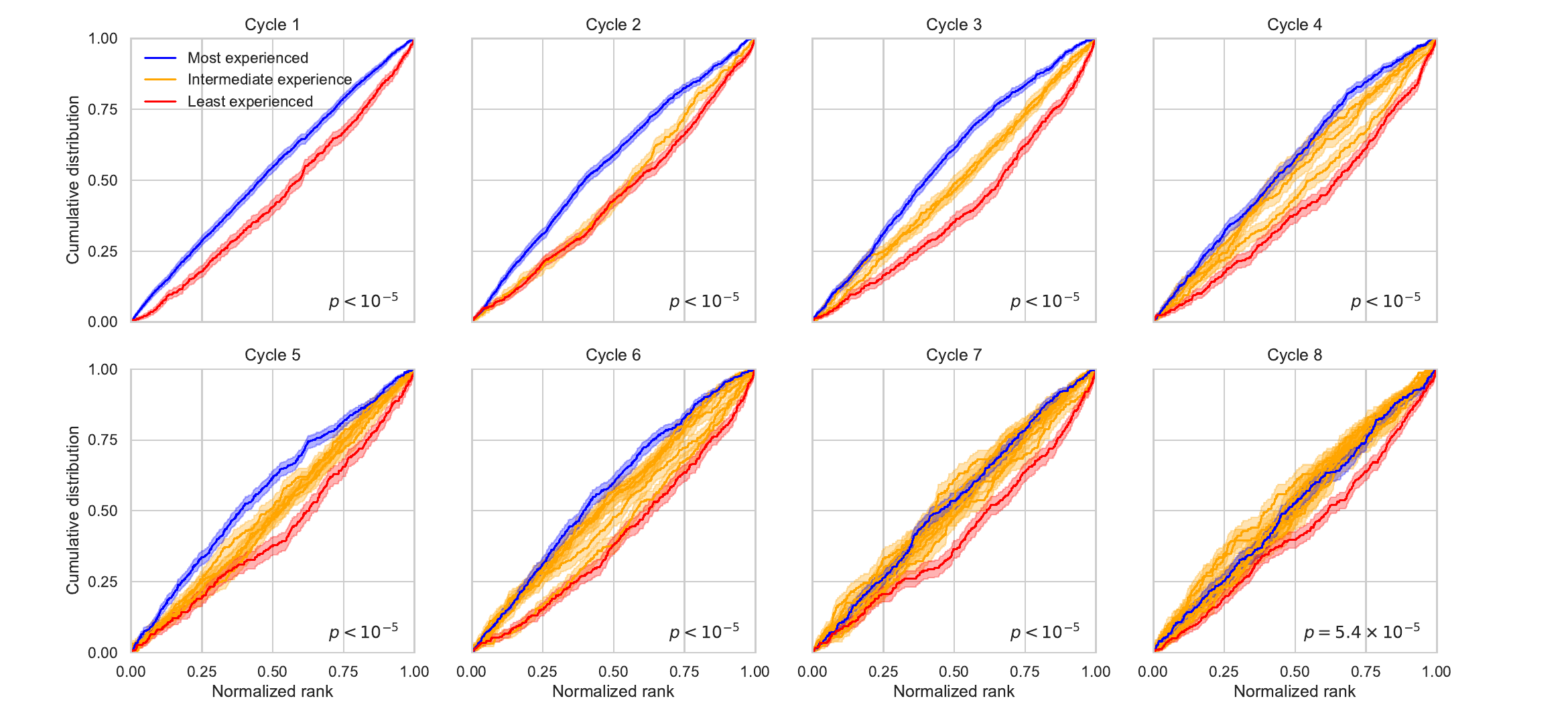}
\caption{
  Normalized cumulative distribution of overall proposal ranks by experience level of the PI. The blue curves show the ranks for PIs who have submitted a proposal in each cycle up to that time. The red curves show the ranks for first-time PIs in that cycle, and the orange curves show intermediate experience levels. The normalized ranks vary between 0 (best) to 1 (worst). The shaded region indicates the 68.3\% confidence interval computed using the beta function. The probability ($p$) that the distributions within a cycle are drawn from the same population, as computed from the Anderson-Darling $k$-sample test, is indicated in the lower right corner of each panel. The results show that in Cycles 0-6, the most experienced PIs had the best overall proposal rankings, but they tended to have more average ranks after randomizing the investigator list (Cycle 7) and introducing dual-anonymous review (Cycle 8). First-time PIs have the poorest overall ranks in each cycle even with the changes introduced in Cycles 7 and 8.
}
\label{fig:ad_experience}
\end{figure}

Figure~\ref{fig:ad_experience} shows the cumulative distribution of proposal ranks by experience level of the PI for all ALMA cycles to date. The ranks are normalized such that a rank of 0 is the best proposal rank and 1 is the poorest. Therefore, distributions shifted to the upper left have better overall proposal ranks than distributions shifted to the lower right. For Cycles 0-7, the Stage 1 ranks, which are set before the panel discussions, are used (see the discussion in Section~\ref{sec:overview} and \citealt{Carpenter20a}). For Cycle 8, the combined distributed peer review and Stage 1 panel results are shown as described in Section~\ref{sec:overview}. In each panel, the blue curve shows the distribution of proposal ranks for PIs who have submitted at least one proposal in each cycle up to that time. The red curves show the ranks for PIs who have submitted a proposal for the first time. The orange curves show intermediate levels of experience. 

Any differences between cumulative distributions were assessed using the Anderson-Darling $k$-sample test \citep{Scholz87} as implemented in {\tt scipy} \citep{Jones01}. The Anderson-Darling test statistic was then used to compute the probability ($p$, $0 < p < 1$)
that the $k$ samples are drawn from the same (but unspecified) population using the {\tt pval} function within the {\tt kSamples} package \citep{Scholz19} designed for {\tt R} \citep{R}. A low value of $p$ suggests that the $k$ samples are drawn from different distributions while a high value of $p$ suggests that the $k$ samples have similar distributions. Any differences in the cumulative ranks are  defined as ``significant" if the probability that the distributions are drawn from the same population is $p<0.01$ and ``marginally significant" if the probability is $0.01\le p \le 0.10$.

\begin{figure}
\centering
\includegraphics[width=\textwidth, clip]{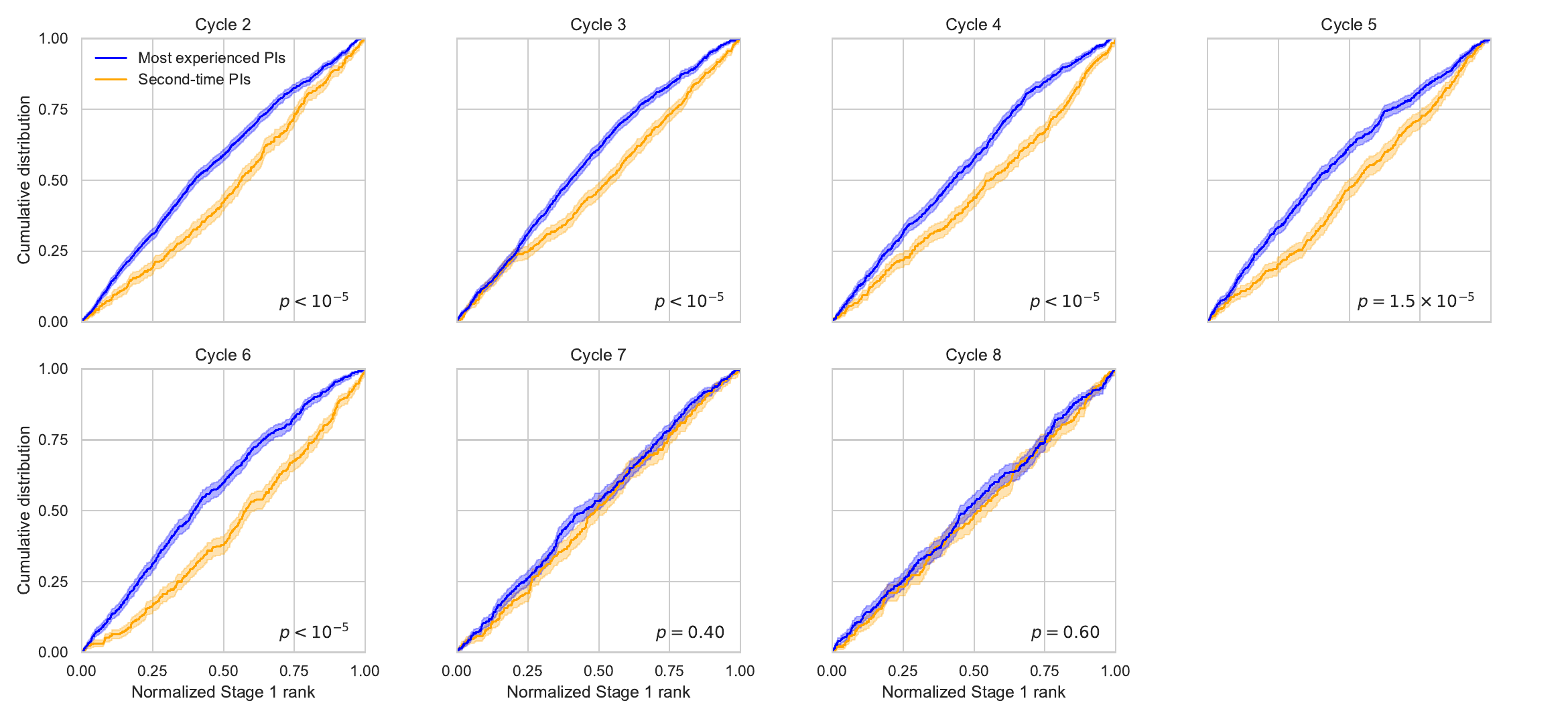}
\caption{
  Normalized cumulative distribution of overall proposal ranks for the most experienced PIs (blue) and second-time PIs (orange) by cycle. Through Cycle 6, second-time PIs always had poorer overall rankings than the most experienced PIs. After randomizing the investigator list (Cycle 7) and implementing dual-anonymous review (Cycle 8), the distributions of ranks are indistinguishable.
}
\label{fig:ad_experience_2}
\end{figure}

As seen in Figure~\ref{fig:ad_experience}, the most experienced PIs no longer have the best overall proposal ranks. Instead, their ranks are consistent with intermediate experience levels. This change first appeared in Cycle 7 when the investigator list was randomized on the proposal cover sheet, and it continued in Cycle 8 when dual-anonymous review was implemented. First-time PIs continued to have the poorest overall ranks in Cycle 7 and Cycle 8. To investigate the impact on intermediate experience levels, Figure~\ref{fig:ad_experience_2} compares the distribution of proposal ranks for the most experienced PIs and PIs who submitted a proposal for the second time. In Cycles 2-6, the most experienced PIs  had better proposal ranks than second-time PIs with high significance. In Cycle 7 and Cycle 8, the two distributions are indistinguishable. This convergence of  ranks is not only a result of the ranks for the most experienced PIs becoming more average (see Figure~\ref{fig:ad_experience}), but also because the ranks of second-time PIs improved.

\begin{figure}
\centering
\includegraphics[width=\textwidth, clip]{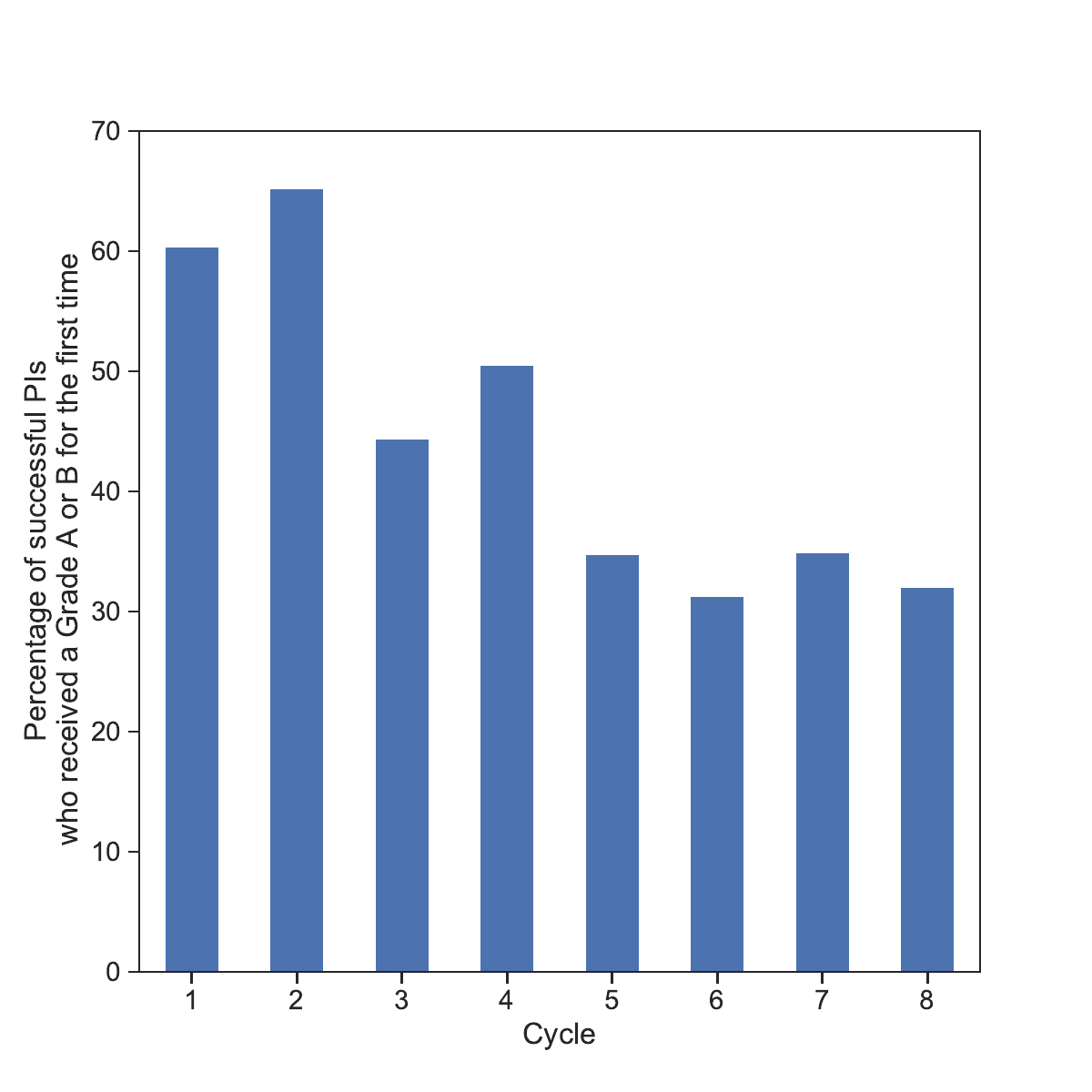}
\caption{
  Percentage of successful ALMA PIs in a given cycle who had a proposal accepted for the first time with priority Grade A or B.
}
\label{fig:success_firsttime}
\end{figure}

Another metric to assess the impact of dual-anonymous review is the fraction of PIs that were successful for the first time in a given cycle. After HST implemented dual-anonymous review, the percentage of PIs that had their HST proposal accepted for the first time increased to $\sim 30\%$ from $\sim 5\%$ in the decade prior \citep{SinghChawla21}. The equivalent analysis for ALMA is presented in Figure~\ref{fig:success_firsttime}, which shows the percentage of PIs by cycle that were awarded a priority Grade A or B proposal for the first time. The total number of hours offered for Grade A and B proposals steadily increased from 700~h in Cycle 0 to 3000 h in Cycle 4, with further increases to 4000 h (Cycles 5 and 6) and 4300 h (Cycles 7 and 8). The percentage of PIs who were successful for the first time was high in Cycles 1-4 when the amount of observing time awarded increased significantly. In Cycles 5-8, where the amount of available time changed by less than 10\%, dual-anonymous review had no obvious impact on the percentage of PIs whose proposal was accepted for the first time. Interestingly, $\sim30\%$ of the ALMA PIs in recent cycles had their proposal accepted for the first time, which is similar to the fraction reported by HST after implementing dual-anonymous review.

\subsection{Regional affiliation}

Following the organization of the ALMA partnership, PIs are grouped into regional affiliations of Chile, East Asia, Europe, North America, or Other. Proposals assigned to East Asia consist of PIs with affiliations in Japan, Taiwan, or the Republic of Korea. Proposals assigned to Europe consist of PIs with affiliations in one of the ESO member states. Proposals assigned to North America consist of PIs from the United States, Canada, or Taiwan. Since Taiwanese agencies contribute funding for ALMA in both East Asia and North America, PIs from Taiwan  can choose to submit a proposal with an East Asian or North American affiliation. For the purpose of this study, these proposals are assigned to East Asia. Proposals from non-ALMA regions are grouped as Other.

\begin{figure}
\centering
\includegraphics[width=\textwidth, clip]{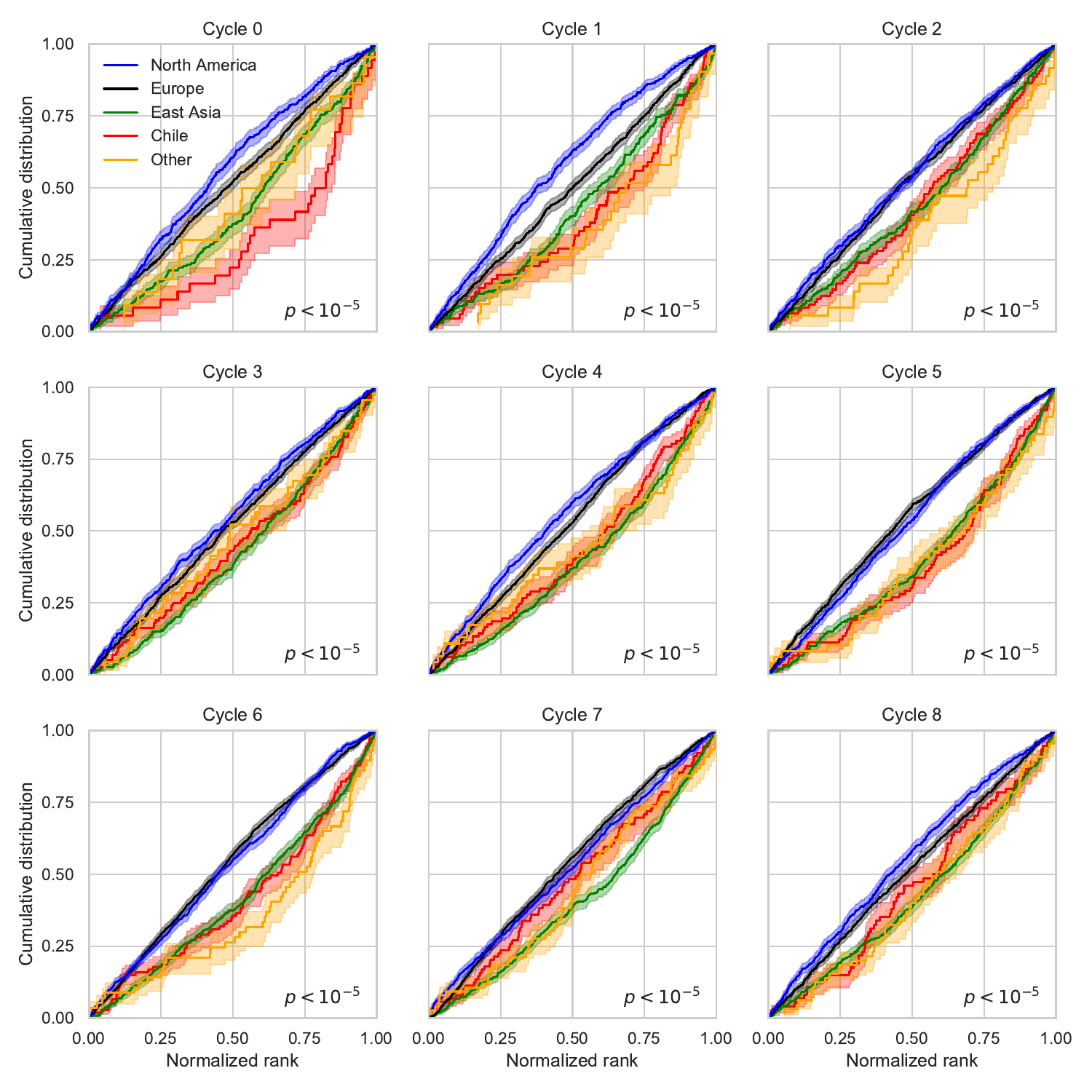}
\caption{
  Normalized cumulative distribution of overall proposal ranks by regional affiliation of the PI. Proposals from Chile, East Asia, and Other regions continue to have poorest proposal ranks relative to PIs from Europe and North America even after randomizing the investigator list (Cycle 7) and the introduction of dual-anonymous review (Cycle 8), although the ranks of Chilean proposals may have improved slightly.
}
\label{fig:ad_executive}
\end{figure}

Figure~\ref{fig:ad_executive} shows the cumulative distribution of proposal ranks by regional affiliation for each ALMA cycle. In Cycles 0-6, PIs from Chile, East Asia, and Other regions all had poorer proposal ranks than PIs from Europe and North America. In both Cycles 7 and 8, these trends continued for East Asia even after randomizing the investigator names and using dual-anonymous review. Evidently the identity of the proposal team is not a significant factor contributing to the below average ranks from East Asia. While the trend of poorer ranks also continued with Chile, the magnitudes of the difference in Cycles 7 and 8 are not as large as in previous cycles.

\begin{figure}
\centering
\includegraphics[width=\textwidth, clip]{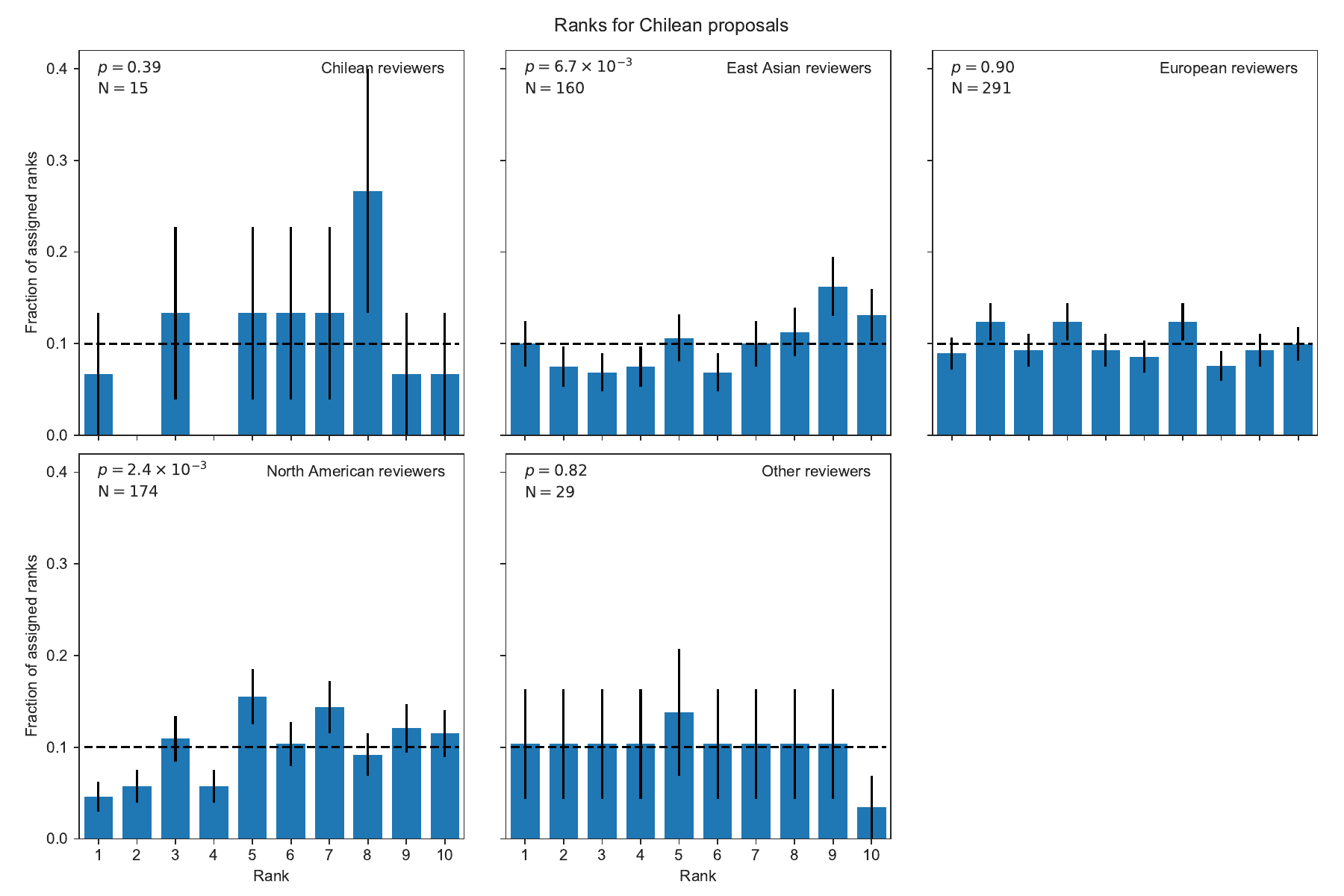}
\caption{
Distribution of ranks assigned to Chilean proposals by reviewers from each region. Each panel is normalized by the number of ranks assigned from the region. The vertical lines in each bin indicate the 1$\sigma$ Poisson uncertainties. The horizontal dashed line shows the expected distribution for uniformly assigned ranks. The number of ranks is indicated by N in each panel, and $p$ indicates the probability that the observed distribution of ranks is consistent with a uniform distribution from the Anderson-Darling test.
}
\label{fig:ranks_cl}
\end{figure}

\begin{figure}
\centering
\includegraphics[width=\textwidth, clip]{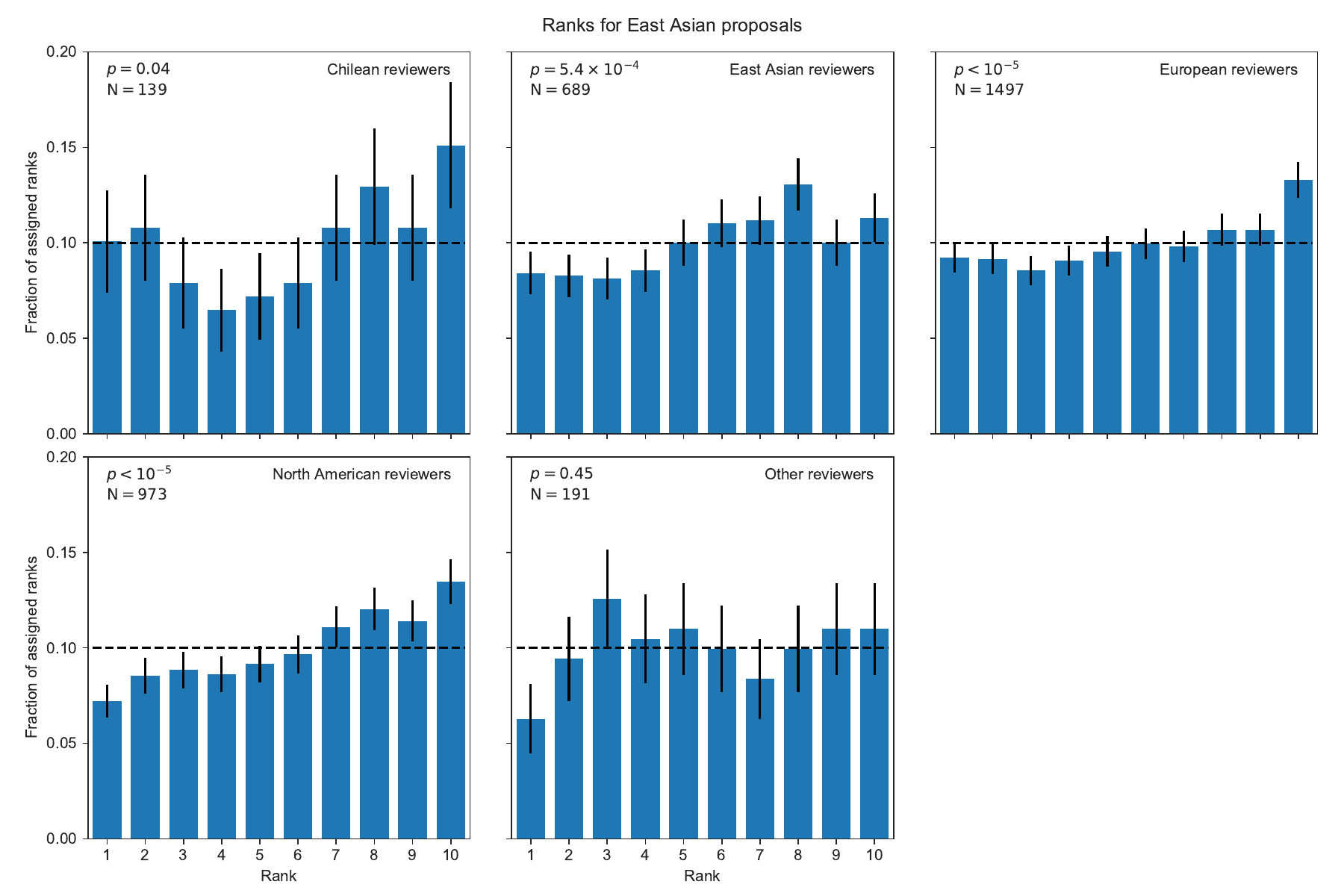}
\caption{
Same as Figure~\ref{fig:ranks_cl}, but for ranks assigned to East Asian proposals.
}
\label{fig:ranks_ea}
\end{figure}

\begin{figure}
\centering
\includegraphics[width=\textwidth, clip]{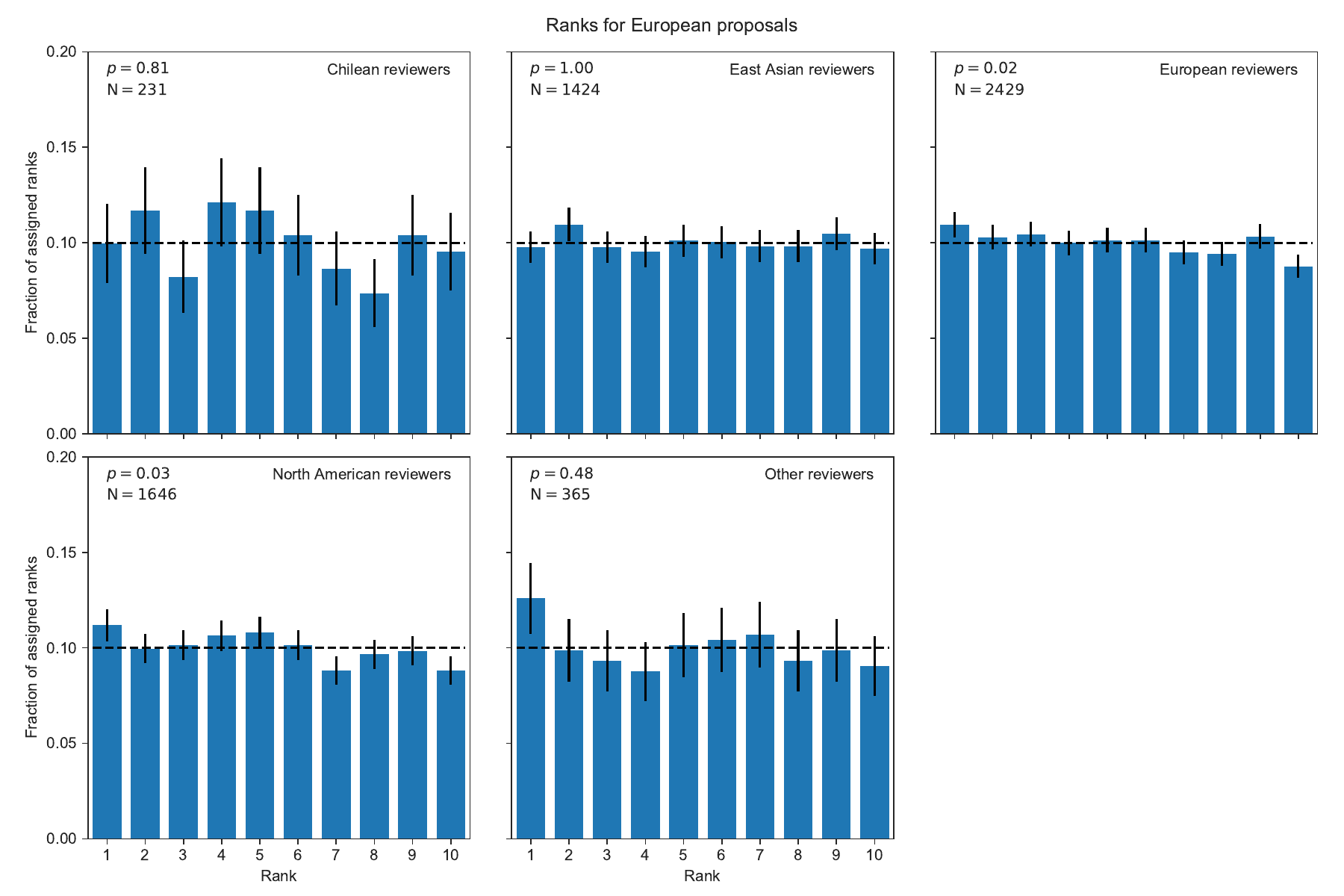}
\caption{
Same as Figure~\ref{fig:ranks_cl}, but for ranks assigned to European proposals.
}
\label{fig:ranks_eu}
\end{figure}

\begin{figure}
\centering
\includegraphics[width=\textwidth, clip]{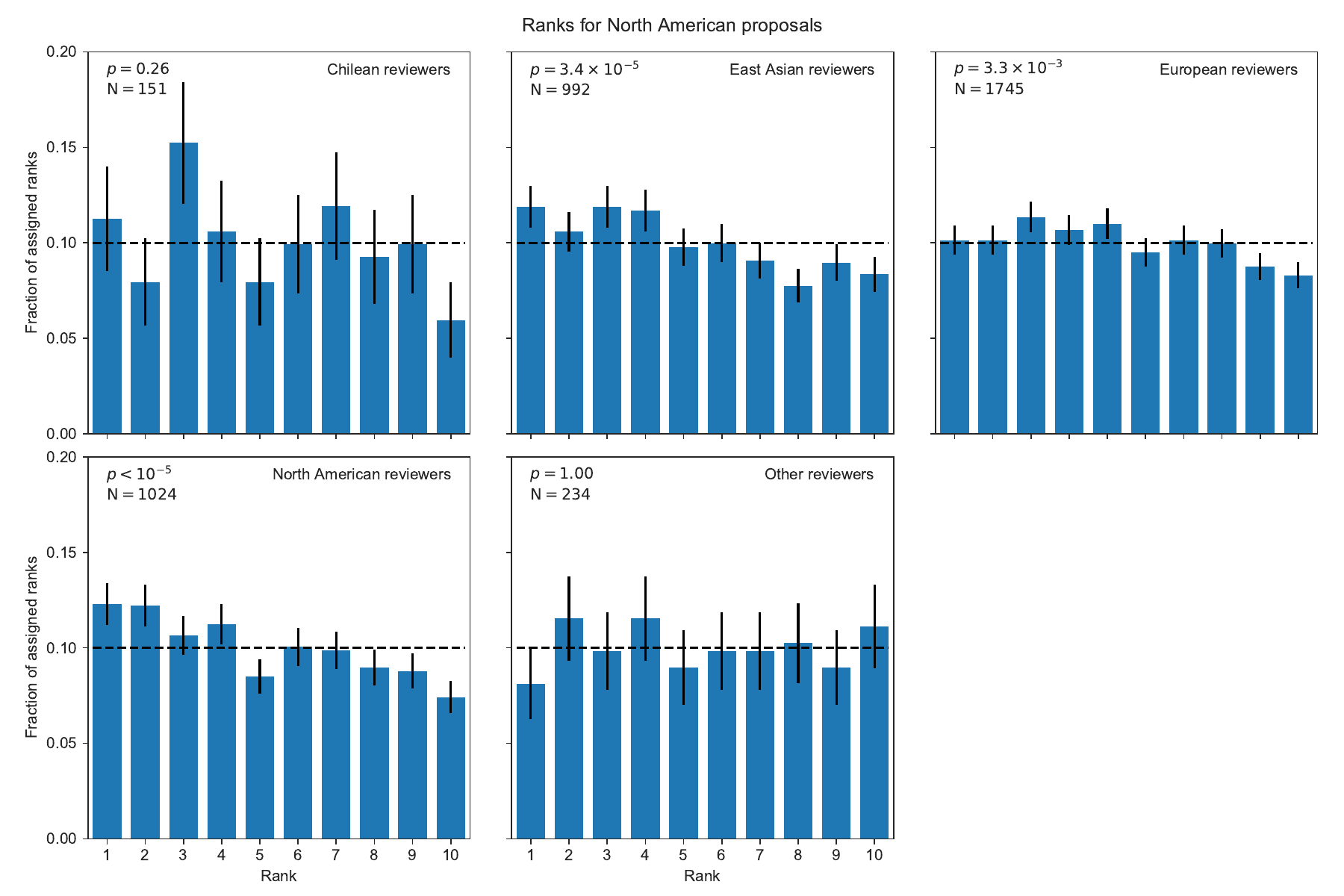}
\caption{
Same as Figure~\ref{fig:ranks_cl}, but for ranks assigned to North American proposals.
}
\label{fig:ranks_na}
\end{figure}

\begin{figure}
\centering
\includegraphics[width=\textwidth, clip]{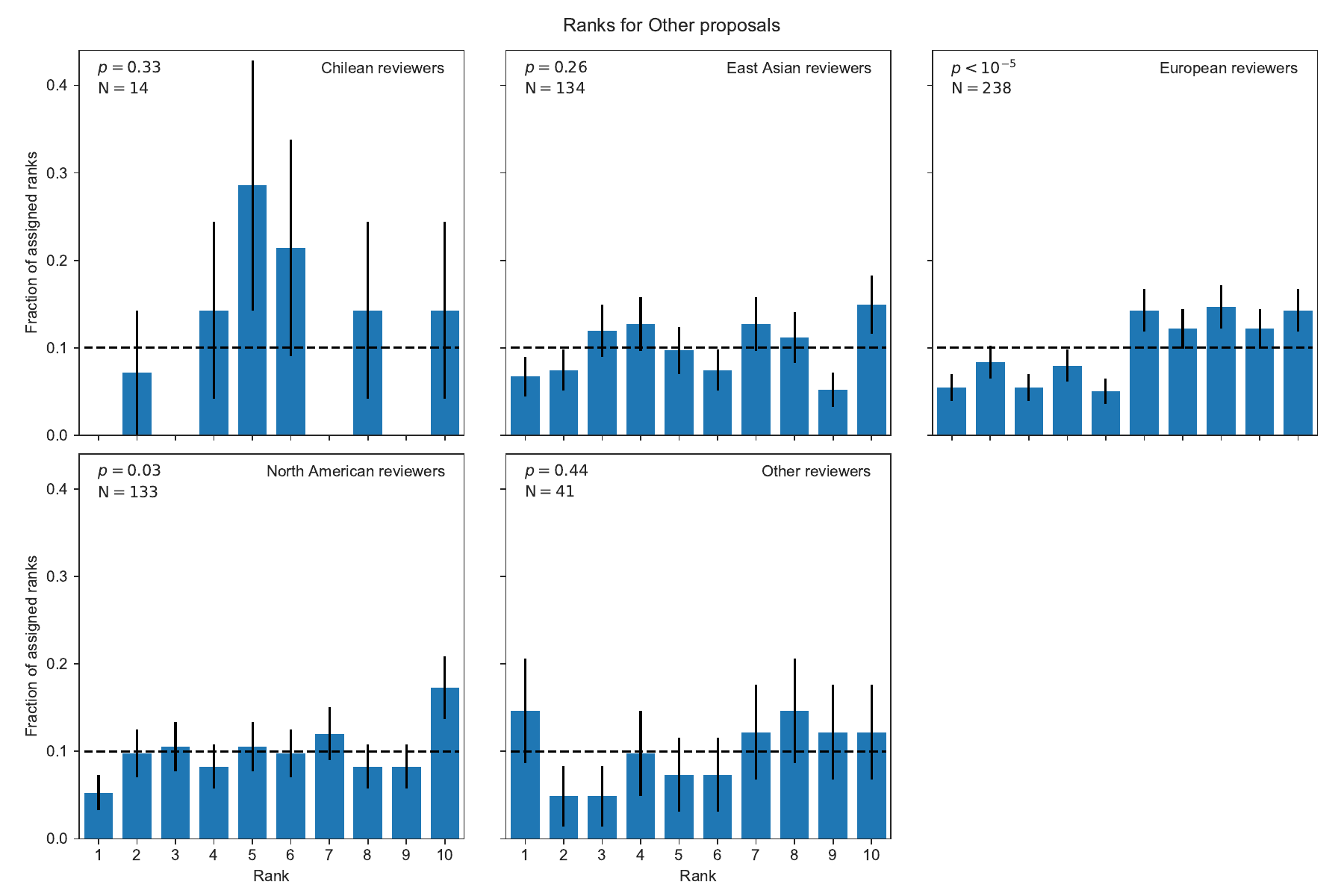}
\caption{
Same as Figure~\ref{fig:ranks_cl}, but for ranks assigned to Other proposals.
}
\label{fig:ranks_other}
\end{figure}

To further investigate the origin of the regional systematics, we compared how the ranks assigned to Cycle 8 proposals that went through distributed peer review depend on the regional affiliation of the reviewers. Figure~\ref{fig:ranks_cl} shows the distribution of ranks assigned to Chilean proposals. Each panel in the figure shows the distribution of ranks assigned to Chilean proposals by reviewers from each region (Chile, East Asia, Europe, North America, and Other). The histograms are normalized by the total number of ranks assigned by the reviewers in a given region, such that the histograms have an expectation value of 0.1 if the ranks are distributed uniformly. The probability ($p$) that the ranks were drawn from a uniform distribution was computed using the Anderson-Darling test and is shown in each panel. Results are also shown for PIs from East Asia (Figure~\ref{fig:ranks_ea}), Europe (Figure~\ref{fig:ranks_eu}), North America (Figure~\ref{fig:ranks_na}), and Other regions (Figure~\ref{fig:ranks_other}).

As shown in Figure~\ref{fig:ranks_cl},  East Asian and North American reviewers tended to give poorer ranks to Chilean proposals. The ranks assigned to Chilean proposals by European reviewers are consistent with a uniform distribution. This is not a result of small number of statistics, since European reviewers had the largest number of proposal assignments. For East Asian proposals (see Figure~\ref{fig:ranks_ea}), the distribution of ranks for reviewers from East Asia, Europe, and North America show significant  departures from a uniform distribution ($p<0.01$). The observed distributions are such that the reviewers assigned more weak ranks (ranks $>$ 5) than strong ranks (ranks $<$ 5). Therefore, reviewers from all three regions tended to assign poorer ranks to East Asian proposals. Both European and North American reviewers tended to rank proposals from Europe better than average with marginal significance ($p=0.02-0.03$; see Figure~\ref{fig:ranks_eu}). Reviewers from East Asia, Europe, and North America all assigned significantly better ranks to proposals from North America (Figure~\ref{fig:ranks_na}).

\begin{figure}
\centering
\includegraphics[width=\textwidth, clip]{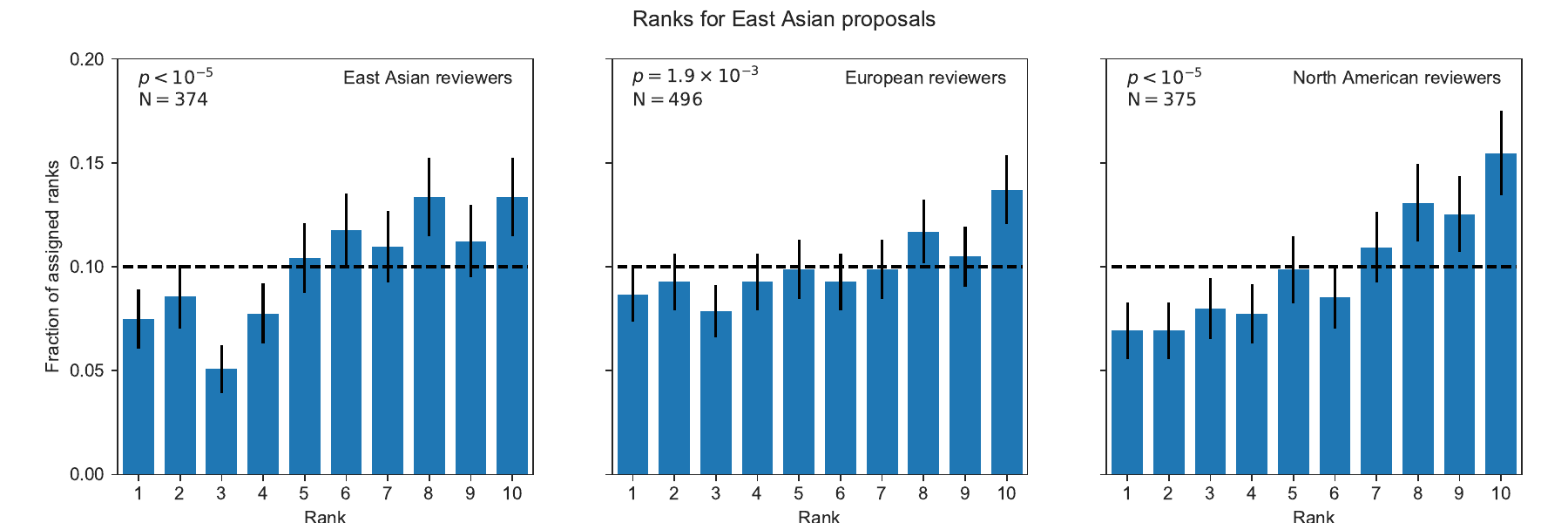}
\caption{
Same as Figure~\ref{fig:ranks_cl}, but for ranks assigned to East Asian proposals in Category 3 from reviewers in East Asia, Europe, and North America. 
}
\label{fig:ranks_c3_ea}
\end{figure}

\begin{figure}
\centering
\includegraphics[width=\textwidth, clip]{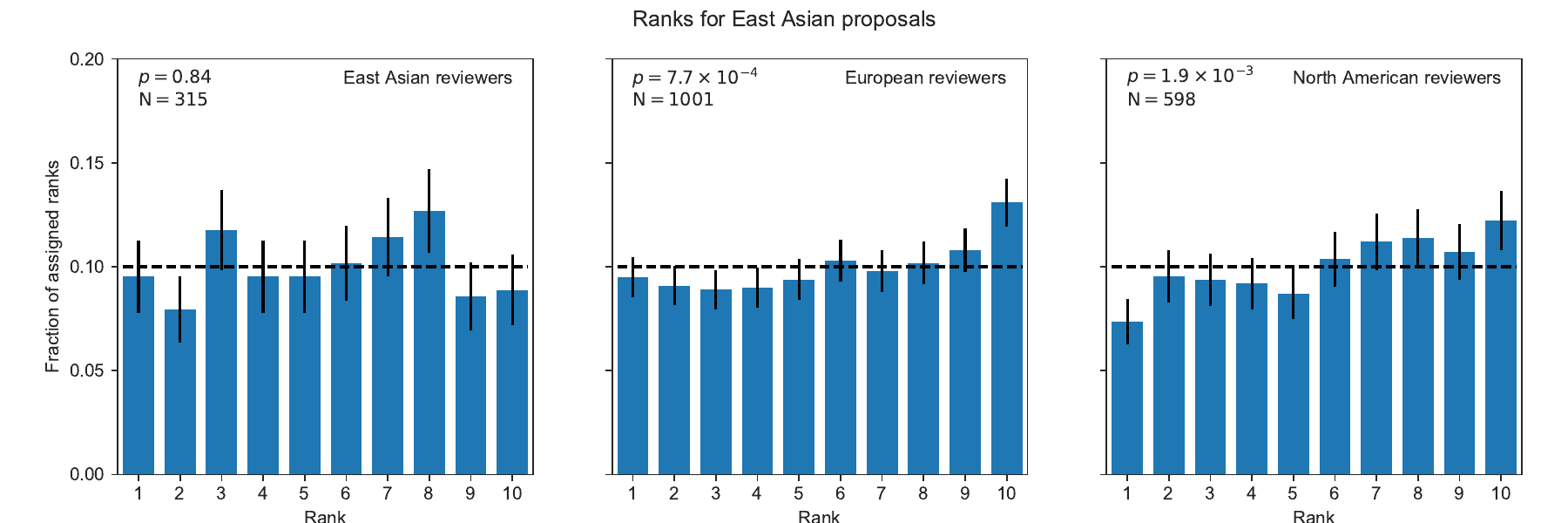}
\caption{
Same as Figure~\ref{fig:ranks_cl}, but for ranks assigned to East Asian proposals in categories 1, 2, 4, and 5 from reviewers in East Asia, Europe, and North America. 
}
\label{fig:ranks_c1245_ea}
\end{figure}

To further explore the origin of the poorer ranks for East Asian proposals, we examined the ranks in different science categories, focusing on reviewers from East Asia, Europe and North America, since these regions have the largest number of reviewers. Figure~\ref{fig:ranks_c3_ea} shows the ranks assigned to East Asian proposals in science Category 3,\footnote{Category 1 is Cosmology and the high redshift universe. Category 2 is Galaxies and galactic nuclei. Category 3 is Interstellar medium, star formation and astrochemistry. Category 4 is Circumstellar disks, exoplanets and the solar system. Category 5 is  Stellar evolution and the Sun.} and  Figure~\ref{fig:ranks_c1245_ea} shows the ranks assigned to East Asian proposals in categories 1, 2, 4, and 5 combined. In both groups of categories, European and North American reviewers tended to give poorer ranks to East Asian proposals. However, for East Asian reviewers, the trend is detectable only in Category 3, while the distribution of ranks in categories 1, 2, 4, and 5 combined are consistent with a uniform distribution. Comparison of the distribution of ranks for East Asian PIs from East Asian reviewers for the two groups of categories also shows a significant difference ($p=0.005$), so the different results between categories is not merely the result of fewer East Asian reviewers in categories 1, 2, 4, and 5 combined versus Category 3.

\subsection{Gender}

The gender of a PI was estimated primarily by using information available on the internet or familiarity with the PI by the authors of this paper or colleagues. In a small number of cases, the gender was gathered from the ALMA user profiles where PIs can volunteer to provide this information. For simplicity, genders have been classified as ``male" or ``female" for this study, but we acknowledge that this is a more complex and nuanced topic in reality. While some estimates of the gender may be in error, we expect the general trends to be correct.

\citet{Carpenter20a} found that proposals led by men tended to have better overall rankings than women in Cycles 0-6. However, any differences in the rankings were not statistically significant in a given cycle with the exception of Cycle 3, when the differences were marginally significant \citep[see also][]{Lonsdale16}. Women also had a lower acceptance rate for their proposals than men in each cycle even after regional, scientific, and experiences levels were considered. 

\begin{figure}
\centering
\includegraphics[width=\textwidth, clip]{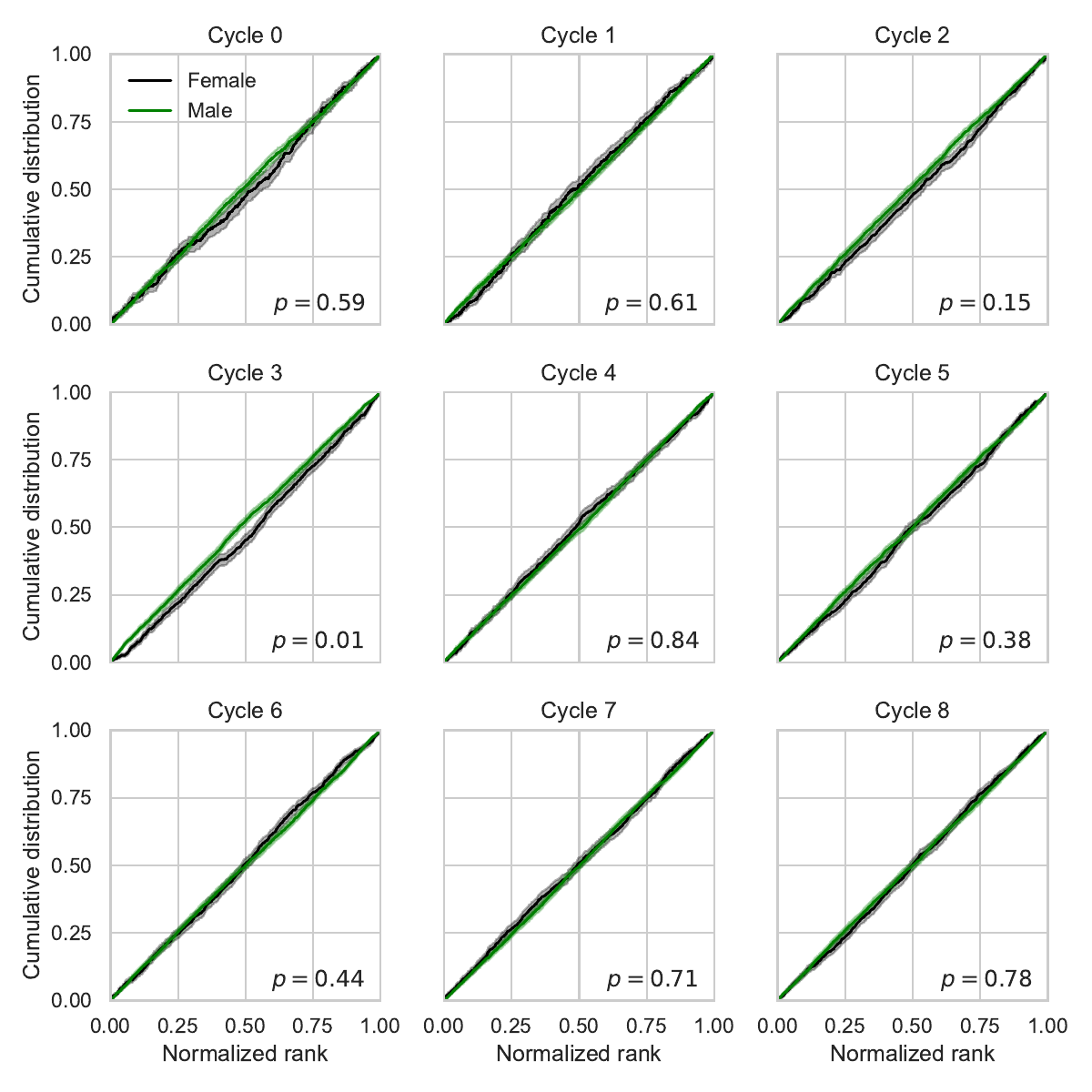}
\caption{
  Normalized cumulative distribution of overall proposal ranks for female (black) and male (green) PIs.
}
\label{fig:ad_gender}
\end{figure}

\begin{figure}
\centering
\includegraphics[width=\textwidth, clip]{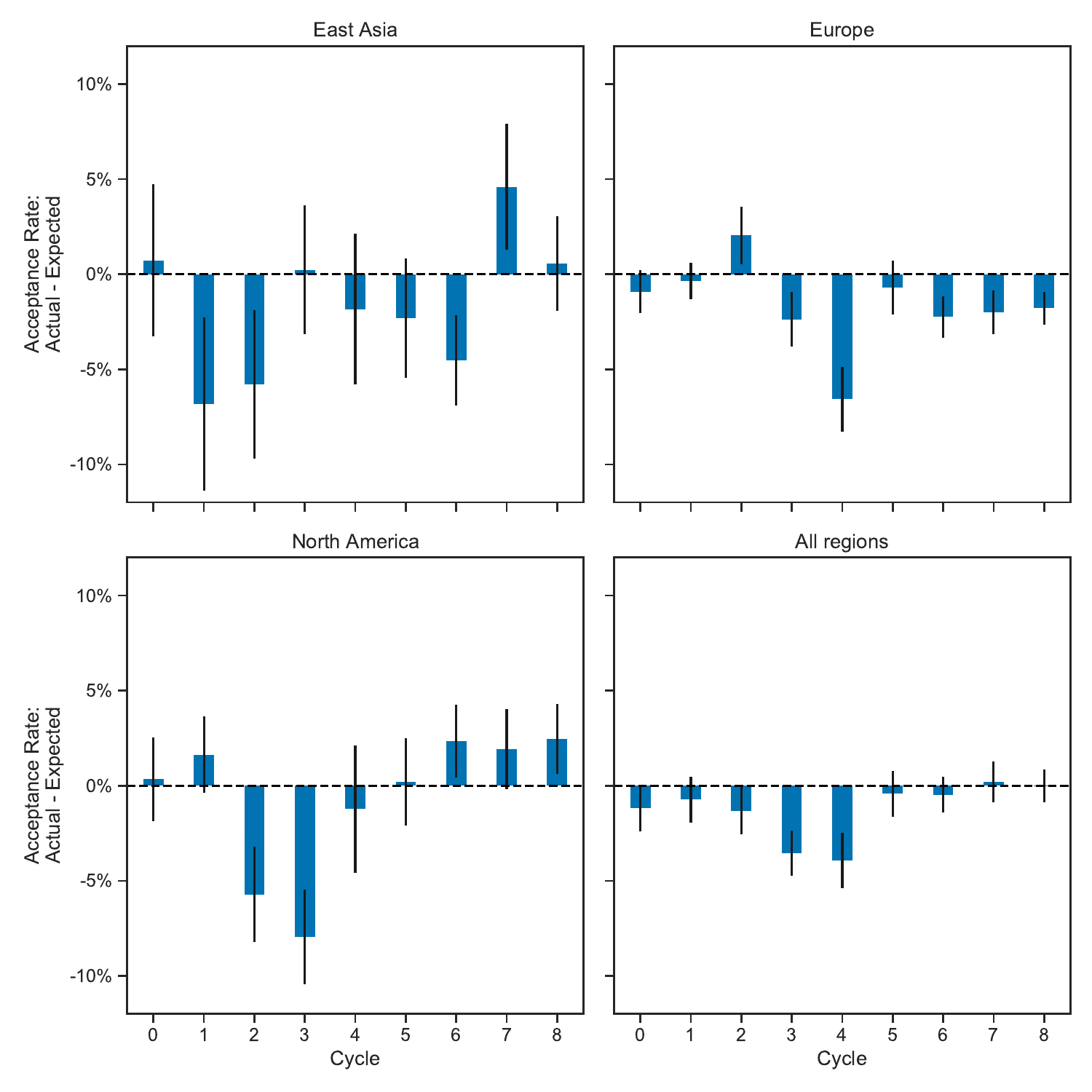}
\caption{
  The difference between the actual and expected acceptance rate of proposals with female PIs by cycle for East Asia, Europe, North America, and all regions combined (including Chile and Other regions). The vertical bars indicate the 1$\sigma$ Poisson uncertainties.
}
\label{fig:acceptance}
\end{figure}

Figure~\ref{fig:ad_gender} shows the cumulative distribution of proposal ranks by gender through Cycle 8. In Cycle 7, women had slightly better ranks than men when measured at the first quartile and slightly poorer ranks in Cycle 8. However, in neither case are the differences significant. Figure~\ref{fig:acceptance} shows the difference between the actual and expected accepted rates for each cycle, where ``accepted" proposals refers to proposals assigned priority grade A or B. As described in \citet{Carpenter20a}, the expected acceptance rate factors the differences in the gender distribution of proposals by region, experience level, and scientific category. In Cycles 0-6, women had a lower acceptance rate than men in each cycle even after considering the demographics. In both Cycles 7 and 8, women had a higher acceptance rate than expected based on the demographic model. However, since Cycle 5, the difference between expected and actual acceptance rates for women is within the 1$\sigma$ uncertainties for all regions combined. The variations in the acceptance rates by cycle are too small to infer any meaningful impact on the distribution of proposal ranks by gender from randomizing the investigator list or implementing dual-anonymous review.

%% file: conclusions.tex
\section{Summary and Conclusions}
\label{sec:conclusions}

We have updated the analysis of the ALMA proposal ranks from Cycles 0-6 presented by \citet{Carpenter20a} to include the results from Cycles 7 and 8. To reduce potential biases in the review processes, the list of investigators on the proposal cover sheet were randomized in Cycle 7 such that the reviewers knew the identity of the proposal teams but not the PI. In Cycle 8, ALMA instituted dual-anonymous review such that the reviewers did not know the identity of the proposal teams.  In addition, Cycle 8 saw the introduction of distributed peer review for the majority of proposals as opposed to the traditional panel reviews used in previous cycles. While distributed peer review was not necessarily expected  to reduce biases in the review process, it is important to examine the results nonetheless for unintended impacts. The ranks were analyzed with respect to the experience level of the PI, the regional affiliation of the PI, and the gender of the PI. 

First-time PIs continue to have the poorest overall proposal ranks even with dual-anonymous review. This suggests that the poor ranks for this demographic group reflect the learning curve in proposing to use a complex instrument such as ALMA. The most significant impact of not revealing the identity of the PI to reviewers is found for PIs who submit proposals in every cycle. In Cycles 0-6, such PIs had the best overall proposal ranks, while in Cycles 7 and 8, these PIs had similar ranks as other PIs (excluding first-time PIs). In addition, PIs submitting a proposal for the {\it second} time now have proposal ranks similar to those of the most experienced PIs, both because the most experienced PIs did not perform as well as they had previously, and the ranks for second-time PIs improved. These results suggest that biases related to name recognition (or lack thereof) have been present in the ALMA review process. Despite the changes seen in the systematics associated with the experience of PIs in submitting ALMA proposals, $\sim$30\% of the successful PIs in Cycle 8 received a Grade A or B for the first time, which is the approximately the same as in recent cycles.

The systematics in the proposal rankings by region continue to persist even after implementing dual-anonymous review. PIs from Chile, East Asia, and Other regions had poorer ranks than PIs from Europe and North America. The differences in the ranks for Chile relative to Europe and North America is somewhat reduced compared to previous cycles, but the difference between East Asia relative to North America and Europe is just as large as in previous cycles. The origin of these systematics was examined in the distributed peer review process by evaluating how reviewers in one region tend to rank proposals in their own or other regions. Reviewers from North America and Europe tend to give better ranks to proposals from those two regions and to give poorer ranks to proposals from East Asia. Reviewers from East Asia also tend to give poorer ranks to East Asian proposals, although this can be attributed primarily to proposals in Category 3 (Interstellar medium, star formation, and astrochemistry). 

\citet{Carpenter20a} speculated on the factors that could contribute to the systematics in the proposal rankings by region, including differences in communication styles between regions and the challenges in communicating in a non-native language. The precise underlying cause remains elusive even with these new data. Given that the trends persist even with dual-anonymous review suggests the systematics are not caused by unconscious biases related to identifying the proposal as being from East Asia using the investigator names. Communication style may still be a factor if PIs from East Asia tend to write proposals in a style different from Europe or North America. Since East Asian reviewers also tend to give poorer ranks to proposals from East Asia, reviewers could prefer reading a particular style versus writing in that style. However, the fact that East Asian reviewers gave poorer ranks to East Asian proposals predominantly in a single category complicates the interpretation. It will be important to see which trends continue in future cycles.

No significant systematics in the proposal ranks are found with respect to gender of the PI in Cycle 8, consistent with the results from the recent cycles. Therefore, randomizing the investigator list and implementing dual-anonymous review have not had a significant impact on the rankings by gender. Nonetheless, for the first time in Cycle 7 and also in Cycle 8, female PIs had a higher acceptance rate than expected just based on  demographics. The difference between the expected and actual acceptance rates has been less than 1$\sigma$ since Cycle 5.

Any changes in the systematics in the past two cycles were largely accomplished by randomizing the investigator list in Cycle 7. No additional significant changes were observed by implementing dual-anonymous review in Cycle 8. Nonetheless, ALMA will continue with dual-anonymous review since the feedback from reviewers and PIs has been largely positive in that they feel it reduces biases in the process and it allows the reviewer to focus on the proposed science.

\software{Astropy \citep{astropy:2018}, SciPy \citep{Jones01}, R \citep{R}, kSamples \citep{Scholz19}}

\begin{acknowledgments}

We would like to thank the referee for their helpful comments on the paper.

\end{acknowledgments}